\pdfoutput=1
\documentclass[11pt]{article}

\usepackage{emnlp2021}

\usepackage{times}
\usepackage{latexsym}

\usepackage[T1]{fontenc}
\usepackage[utf8]{inputenc}

\usepackage{microtype}

\usepackage{inconsolata}

\usepackage{adjustbox}
\usepackage{forest}
\usepackage{microtype}
\usepackage{pifont}% http://ctan.org/pkg/pifont
\usepackage{latexsym}

\usepackage{balance}
\usepackage{helvet}  %Required
\usepackage{courier}  %Required
\usepackage{url}  %Required
\usepackage{graphicx}  %Required
\usepackage{amssymb}
\usepackage{amsmath}
\usepackage{algorithm}
\usepackage{algorithmic}
\usepackage{amsthm}
\usepackage{multirow}
\usepackage{pgffor}
\usepackage{epstopdf}
\usepackage{tikz}
\usepackage{epstopdf}
\usepackage{mathtools}
\usepackage{enumitem}
\usepackage{subfigure}
\usepackage{tabulary}
\usepackage{color}
\usepackage{url}
\usepackage{flushend}
\usepackage{stfloats}
\usepackage{tcolorbox}
\usepackage[utf8]{inputenc}
\usepackage[english]{babel}
\usepackage{tabularx}
\usepackage{CJKutf8}
\usepackage{microtype}
\usepackage{textcomp}
\usepackage{booktabs}
\usepackage{colortbl}
\usepackage{soul}
\usepackage{xcolor}
\usepackage{arydshln}
\usepackage{natbib}
\usepackage{appendix}

\title{Modeling Uncertainty and Using Post-fusion as Fallback Improves Retrieval Augmented Generation with LLMs}

\author{
  \textbf{Ye Liu}, \textbf{Semih Yavuz}, \textbf{Rui Meng}, \textbf{Meghana Moorthy Bhat},\\ \textbf{Shafiq Joty}, \textbf{Caiming Xiong}, \textbf{Yingbo Zhou} \\
 Salesforce Research\\
  { \texttt{\{yeliu, syavuz, ruimeng, meghana.bhat},} \\ {\texttt{sjoty, cxiong, yingbo.zhou\}@salesforce.com}}
}

\begin{document}
\maketitle
\begin{abstract}
The integration of retrieved passages and large language models (LLMs), such as ChatGPTs, has significantly contributed to improving open-domain question answering. However, there is still a lack of exploration regarding the optimal approach for incorporating retrieved passages into the answer generation process. This paper aims to fill this gap by investigating different methods of combining retrieved passages with LLMs to enhance answer generation. We begin by examining the limitations of a commonly-used concatenation approach. Surprisingly, this approach often results in generating ``unknown'' outputs, even when the correct document is among the top-$k$ retrieved passages.
To address this issue, we explore four alternative strategies for integrating the retrieved passages with the LLMs. These strategies include two single-round methods that utilize chain-of-thought reasoning and two multi-round strategies that incorporate feedback loops. Through comprehensive analyses and experiments, we provide insightful observations on how to effectively leverage retrieved passages to enhance the answer generation capability of LLMs. On three open-domain question answering datesets, NQ, TriviaQA and SQuAD, our multi-round approaches outperform traditional concatenation approach, achieving over a $10\%$ improvement in answer EM.

%We first analyze the limitations of a simple concatenation approach, which we find generates ``unknown'' even when the gold document is among the top-$k$ retrieved passages. Subsequently, we delve into four alternative methods to amalgamate the retrieved passages with the LLM. These consist of two approaches leveraging chain-of-thought reasoning and another two multi-round strategies that employ feedback. Through comprehensive analysis and experimentation, we provide valuable insights into leveraging retrieved passages to enhance LLM's answer ability.
\end{abstract}

% the challenges of using separate document inputs with LLMs, where the correct answer may not be the majority output.

\section{Introduction}

% Large Language Models (LLMs) have powered a range of applications but struggle with underrepresented knowledge in their training data, leading to non-existent information, particularly in open-domain settings. This issue extends to open-domain question answering, a key focus in NLP with numerous applications. Despite the advances made by transformer-based models like GPT~\cite{brown2020language,bubeck2023sparks}, these models often falter in generating precise responses in a fully unsupervised setting, especially when answers necessitate specific facts or intricate details.

Large Language Models (LLMs), such as GPTs~\cite{brown2020language,bubeck2023sparks}, have found extensive applications, but often struggle with limited knowledge representation, resulting in inaccuracies and insufficient specificity in open-domain question answering. To overcome these limitations, the integration of retrieval-based techniques~\cite{izacard2022few,borgeaud2022improving} has emerged as a promising solution. By incorporating relevant passages during the answer generation, LLMs can leverage external information to provide more accurate and detailed responses. Nevertheless, effective strategies for incorporating retrieved passages into the LLMs remains a challenging and relatively understudied area. 

\begin{figure}[t]
\centering
\includegraphics[width=\linewidth]{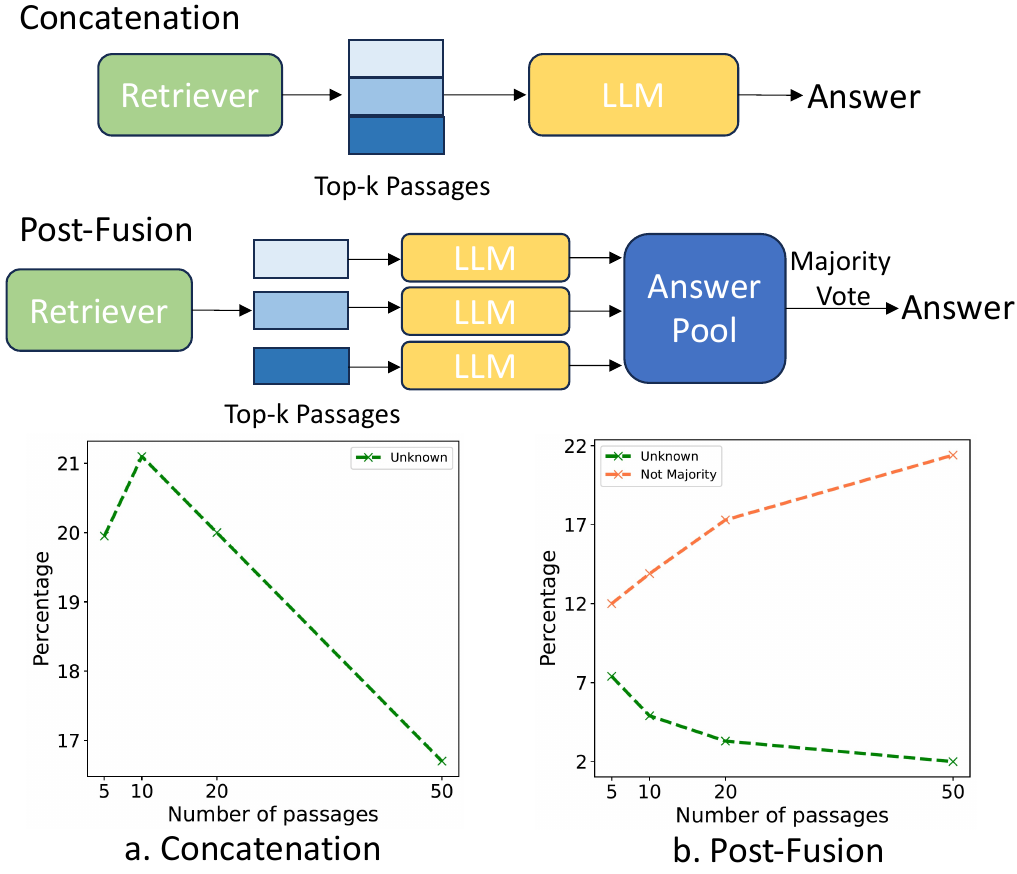}
\caption{\textbf{Top}: Illustration of Concatenation v.s. Post-Fusion strategies. \textbf{Bottom-a}: percentage of unknown responses using the Concatenation strategy. \textbf{Bottom-b}: by varying the number of retrieved passages, (\textcolor{teal}{green}) percentage of unknown responses, and (\textcolor{red}{red}) error rate by majority voting (when the correct answer is in the answer pool, the majority selects a wrong answer).}
\label{figure:intro}
\end{figure}

Our analysis (Fig.~\ref{figure:intro}), conducted under the oracle setting where one of the top-$k$ retrieved passages contains the answer, reveals that a simple concatenation of the passages into LLMs often leads to ``unknown'' responses — instances where the provided context fails to answer the question — accounting for about 20\% of all responses. An alternative method, where the passages are individually provided as input to LLMs and the majority vote determines the final answer, reduces the rate of ``unknown'' generation to 2-7\% depending on the number of passages. However, this method introduces a new challenge: the correct answer does not align with the majority vote in the answer pool. Particularly, when more passages are incorporated from 5 to 50, the error rate of the majority vote increases from 12\% to 22\%. Thus, both of the methods have their own weaknesses and more suitable approaches for the integration of retrieved passages and LLMs remain to be investigated.

Transformer-based LLMs have shown the capability to utilize attention mechanisms~\cite{vaswani2017attention} for discovering token-level relevance. However, they may not always attend to the relevant parts within the context, leading to a potential oversight of important information present in the retrieved passages~\cite{clark2020electra,zhao2020transformer}. This challenge becomes more pronounced when dealing with extensive corpora like Wikipedia, which contains over 21 million passages, making it a formidable task to identify the most relevant passages for a question. Furthermore, retrieved passages that are closely related to the question's topic can act as distractors, potentially misleading the model~\cite{asai2019learning}. If the model mistakenly directs its attention towards these distractor passages, it can introduce noise that negatively impacts the answer prediction process.

In this paper, we explore the integration of retrieved passages with LLMs like ChatGPTs to enhance their ability to generate correct answers. In particular, we examine situations where the retrieved passages contain the correct answer, yet the model fails to generate the correct response, indicating areas for improvement. Initially, we focus on two chain-of-thought (CoT) \cite{wei2022chain,wang2022self,trivedi2022interleaving} strategies that incorporate in-context learning. We introduce a pruning strategy and a summarization strategy for the retrieved passages to guide the answer generation process of the LLMs. 

Subsequently, we investigate two multi-round methods with feedback: \textbf{Post-Fusion as the Fallback}: In the initial round, this method employs the Concatenation approach to generate an answer. If the LLM generates ``unknown'' responses with the inputs, it proceeds to use Post-Fusion in the second round, generating candidate answers. The final answer is chosen via majority vote.  
\textbf{Concatenation as the Distiller}: This approach starts by leveraging Post-Fusion to produce a pool of potential answers and to identify relevant passages. In the subsequent round, only the unfiltered passage is concatenated with the question and answer candidates from the first round. This consolidated input is then fed into the LLM to derive the final answer.  

Through extensive experiments on three single-hop open-domain question-answering datasets, we showcase the enhanced performance of our proposed methods, achieved with a minimal additional resource cost. Our findings provide a foundation for the development of more advanced retrieval-integration methods aimed at further enhancing the capabilities of these models. 

%We aim to go beyond the simple concatenation of the question with the retrieved document and instead propose new techniques that may lead to more coherent and factually accurate responses.

% Many chain-of-thought works has been done on multi-hop question answer that typically entails decomposing complex questions into sequential intermediate steps (a.k.a. chains) and in each of the decomposed single-hop question, the 
% The assumption that 
% Retrieval and LLM 

\section{Problem Setup}
This study focuses on the question answering task under the open-domain setting. It remains a open problem to retrieve the most relevant context for question answering. Therefore, a common practice is to include multiple top ranked passages, which serves as the supplementary context for the LLMs. The number of supplementary passages, denoted as $k$, can vary based on the desired input length $M$ of the LLM. Typically, $k$ can be set to 5, 10, or 20, ensuring that the total length of $k$ passages, each having a maximum length of $L$, remains within the maximum input length $M$ of the LLM (i.e., $k*L < M$). By incorporating these supplementary passages, the LLM is provided with a more comprehensive and informative context, which has the potential to enhance its accuracy.

% Based on our analysis, we find that different approaches could influence the answer performance a lot. Therefor, in this study, we want to investigate how to put those $k$ passages into the LLM.

\section{Methods}
We adopt a two-stage pipeline for open-domain QA. It consists of two black-box components, a retriever and a LLM such as ChatGPT and LLama2~\cite{touvron2023llama}. We aim to methodically investigate the optimal methods for transferring the top-$k$ retrieval results to the LLMs for generating factoid answers. Our investigation begins with a focus on various \textbf{single-round} strategies, wherein the retrieved passages are directly fed into the LLMs. Subsequently, we delve into several \textbf{multi-round} approaches, involving the initial supply of retrieved passages to the LLMs, gathering feedback, and then modifying the interaction process with the LLMs based on that feedback.

\subsection{Definition of Unknown Output}
LLMs are not universally capable. Their effectiveness relies on being trained on relevant data, storing essential knowledge within their weights. When an LLM cannot provide an answer directly, a common strategy is to use retrieval to fetch pertinent context. However, there may be instances where the model discerns that the retrieved context is insufficient for a response. In such cases, the LLM might produce outputs like ``The provided input does not contain the context to answer the question.'' We interpret this behavior as the LLM's self-awareness of its inability to confidently produce an answer based on the top-$k$ retrieved passages. To standardize the model's response in these situations and prevent varied output formats, we prompt the model to generate ``unknown'' when it believes the given context is inadequate for an answer. To be specific, we add the following sentence in the prompt: ``\textit{If don't know the answer, just say Unknown.}''

\subsection{Single-Round Approaches}
In this section, we explore single-round strategies where retrieved passages are directly sent to the LLM. We first examine a zero-shot approach, providing only the task definition and desired output format, without demo examples. Then, we study a one-shot strategy, utilizing a single demo example to guide the LLM's answer generation. 

\subsubsection{Zero-shot Prompt}
Our first line of investigation pertains to a zero-shot setting. In this setting, we only provide the task definition and the desired answer format as the prompt, excluding any demonstration examples that elucidate how to generate an answer from the question and the Top-$k$ passages.

\textbf{Concatenation Prompt.}
We begin our exploration with a straightforward and commonly used method that involves concatenating the question and the retrieved passages. These passages are arranged in the order they were retrieved and combined into a single text string. This composite text is then fed into the language model to generate the final answer, which can be represented by the below equation:
\begin{equation}
    a = \mathrm{LLM}(q, p_1, p_2, ..., p_k)
\label{eq_concat}
\end{equation}

From our experimental results, we observe that this approach can potentially lead to ``unknown'' output, even when one of the retrieved passages contains the ideal context necessary to answer the question. This stems from the LLM possibly becoming confused due to the complexity or abundance of input, subsequently generating an unsatisfactory response.

\textbf{Post-Fusion Prompt.}
We also explored an alternative approach where each of the Top-$k$ retrieved passages is independently fed to the LLM. After generating an answer for every passage, the collective responses form an answer pool. A majority voting mechanism is then applied to this pool to determine the final answer, which can be denoted by the following equation:
\begin{equation}
\begin{split}
    a_1 = &\mathrm{LLM}(q, p_1), \cdots ,a_k =\mathrm{LLM}(q, p_k) \\
    &\text{majority} = \arg\max_{i} a_i
\end{split}
\label{eq_separate}
\end{equation}

Our experimental findings suggest that while this approach can decrease the likelihood of indeterminate output, it presents a distinct challenge. Specifically, the correct or ``gold'' answer may indeed be presented within the generated answer pool, but it might not be the majority answer, thus resulting in an incorrect final response.

\subsubsection{Few-shot Prompt}
We introduce two distinct prompts, with one-shot example, to guide the LLMs in fusing answers from potentially relevant passages. Examples of these two prompt types are provided in Fig.~\ref{figure:pruning} and \ref{figure:summary} in the Appendix~\ref{prompt_approach}, respectively.

Given the significant enhancements chain-of-thought brings to multi-hop question answering, we aim to adapt this approach for single-hop retrieval-augmented generation. Our method uses demonstrative examples to guide answer generation strategies.
We employ two techniques for this: One approach involves pruning irrelevant passages and using the few remaining relevant ones for answer generation.
The other one is to initially identify the relevant information and then summarize the relevant information like chain of thought and generate the final answer. 

\textbf{Pruning Prompt.}
This prompt requires the LLM to effectively identify answerable passages through a process of selective elimination. As a result, The demonstration involves differentiating irrelevant passages from the ones that can provide an answer, and subsequently generating the final response based on the few relevant passages. 

\textbf{Summary Prompt.}
Summarization represents a strategy that extracts the central information from the Top-$k$ passages. Based on this synthesized summary, the LLM can produce the final answer. We posit that summarization could serve as a guiding mechanism for the LLM to more effectively respond to the question. To illustrate this, we provide a demonstration example that exhibits how the model selects useful information from the passage before delivering the final response.

%Summarization strategy condenses core information from the top-$k$ passages to guide LLMs in formulating responses~\cite{demner2006answer}. A demonstration example illustrates this process is provided, showcasing how the model should extract useful information before generating answers.

\subsection{Multi-Round Approaches}
In our exploration of multi-round strategies, we first provide the retrieved passages to the LLM. Based on the initial feedback received either ``unknown'' or a list of candidate answers, we adjust our interaction process with the LLM accordingly.

\begin{figure}[t]
\centering
\includegraphics[width=0.95\linewidth]{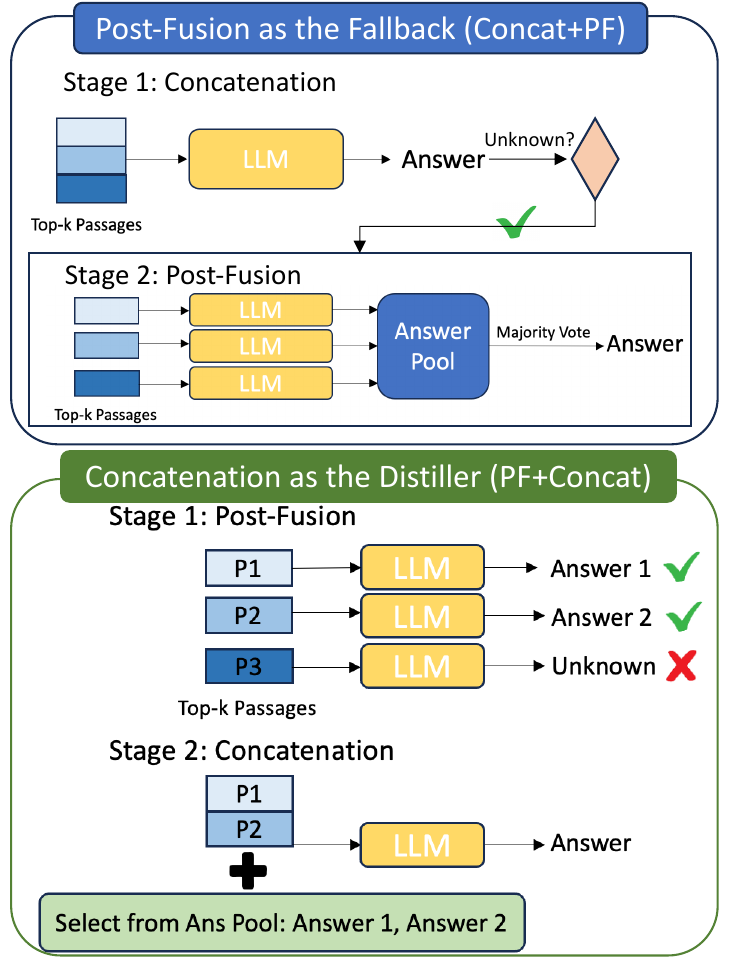}
\caption{Diagram of Post-Fusion as the Fallback on top and Concatenation as the Distiller at bottom.}
\label{figure:concat+pf}
\end{figure}

% \begin{Fig.}[t]
% \centering
% \includegraphics[width=0.7\linewidth]{Fig./PF+Concat.pdf}
% \caption{The token usage of different approaches using Top-5 passages.}
% \label{figure:pf+concat}
% \vspace{-1em}
% \end{Fig.}

\textbf{Post-Fusion as the Fallback (Concat+PF).} %\semih{<-- name suggestion: ""}
Initially, we employ the concatenation method as illustrated in upper box of Fig. \ref{figure:concat+pf} to obtain an answer predicted by the LLM. If the LLM determines that the input passages are unable to provide an answer to the question (i.e., ``unknown'' responses), we then proceed to the second round where we utilize the Post-Fusion approach to produce an answer pool. Finally, we employ a majority vote to select the final answer. 

\textbf{Concatenation as the Distiller (PF+Concat).}
To begin with, we leverage the Post-Fusion strategy to curate a pool of potential answers shown in lower box of Fig. \ref{figure:concat+pf}. Instead of performing a majority vote, a passage selection process~\cite{lewis2020retrieval} is adopted to discard passages that yield an ``unknown'' output by the LLM. In the second round, the LLM is prompted with the concatenation of the unfiltered passages, along with the question and answer candidates generated from the first round. The purpose is to guide the LLM in effectively extracting (distilling) the correct answer from the pool of candidates. 

% To begin with, we leverage the `post-fusion' prompting strategy to curate a pool of potential answers in the first round. \semih{The following sentence is breaking the flow?} In the single-passage question answering phase, we also implement a passage selection process~\cite{lewis2020retrieval} to discard any passages that yield indeterminate or `unknown' output from LLM. Following this, we apply the `concatenation strategy', combining only those passages that can contribute to generating an answer with the potential answer pool. \shafiq{Clarify the following sentence. What is collective input here?} This collective input is then presented to the LLM, which is responsible for choosing the final answer from the assembled candidate answer pool.

\section{Experiments}

\textbf{Evaluation Benchmarks.} We conduct evaluations on multiple datasets of open-domain question answering to assess the performance of the proposed integration approaches. 

The datasets used include Natural Questions (NQ)~\cite{kwiatkowski2019natural}, TriviaQA~\cite{trivedi2022musique}, and SQuAD-Open~\cite{ho2020constructing} are all datasets designed for training and evaluating single-hop question answering models. NQ is sourced from Google Search queries and their corresponding Wikipedia answers. TriviaQA offers a broader domain with trivia questions and their answers derived from web and Wikipedia sources. Conversely, SQuAD-Open is a variant of the original SQuAD dataset that requires the model to extract answers from open-domain Wikipedia content, without any pre-specified passage.

\textbf{Evaluation Metrics}
We adhere to traditional QA dataset evaluation methods~\cite{yang2018hotpotqa,ho2020constructing}, contrasting with the recent LLM evaluations on QA tasks detailed in~\cite{liu2023lost}, which assess whether the generated answer includes the ground truth. Importantly, our evaluation criteria are more rigorous than these recent LLM evaluations~\cite{liu2023lost}, given that we mandate the LLM to adhere strictly to the given prompt in generating an entity-specific answer. In detail, predicted answers are evaluated with the standard answer exact match (EM) and F1 metric~\cite{rajpurkar2016squad,liu2022uni}. A generated response is considered correct if, after normalization, it matches any candidate in a list of acceptable answers. The normalization process entails converting the text to lowercase and omitting articles, punctuation, and redundant whitespaces. 

We also evaluate the percentage of ``unknown'' responses (\%Unk) which gauges the proportion of times the LLM indicates it cannot answer based on the given input. Additionally, we measure the error rate through majority vote (\%NM), representing instances where the correct answer is within the generated answer list but isn't the majority selection.

\textbf{Dataset Filter}
To mitigate the influence of specific training datasets on the LLM~\cite{aiyappa2023can}, we initially prompt the LLM to answer questions without any provided context. This process enables us to filter out questions that the LLM can accurately answer independently, thereby eliminating the need for additional external contextual information. The remaining questions, which the LLM couldn't answer independently, are the focus of our study. This filtering ensures our evaluation stringently reflects the LLM's ability to utilize external context from retrieved passages.

We use the development set of NQ, TriviaQA, and SQuAD, initially containing 5,892, 6,760, 5,928 questions, respectively. After removing questions that can be answered without context, we are left with 3,459 questions in NQ, 1,259 in TriviaQA, and 3,448 in SQuAD.

\begin{table*}[ht]
\centering
\resizebox{0.9\linewidth}{!}{
\renewcommand{\arraystretch}{1}
\begin{tabular}{l|cccc|cccc|cccc}
\toprule
                & \multicolumn{4}{c}{\textbf{NQ}}           & \multicolumn{4}{c}{\textbf{TriviaQA}}     & \multicolumn{4}{c}{\textbf{SQuAD}}          \\ 
                & EM & F1 & \%Unk & \%NM & EM & F1 & \%Unk & \%NM & EM    & F1 & \%Unk & \%NM \\ \hline
 \multicolumn{9}{l}{\textit{\textbf{With gold passage}}}\\ \hdashline
 \texttt{LLama2} \\
Concatenation    & 26.9 & 36.9 & 12.9\% & -     & 38.5 & 44.9 & 8.3\%  &  -     & 37.0 & 39.3 & 10.8\% &  -     \\
Post-Fusion     & 27.5 & 38.6 & 2.8\%  & 27.8\% & 38.8 & 45.2 & 4.4\%  & 19.2\% & 38.3 & 42.3 & 6.8\%  & 8.9\%       \\ 
Pruning Prompt      & 27.8 & 37.8 & 10.9\% & -     & 39.3 & 45.9 & 7.8\%  &  -     & 35.3 & 41.7 & 8.4\%  & -         \\
Summary Prompt    &  28.1 & 37.9 & 9.8\%  & -      & 39.2 & 45.2 & 7.5\%  & -      & 38.5 & 42.6 & 7.9\%  & - \\
Concat + PF & \textbf{30.3} & \textbf{40.5} & \textbf{1.7\%}  & 3.8\%  & 40.4 & 46.0 & \textbf{0.8\%}  & 2.6\%  & \textbf{41.5} & \textbf{45.1} & \textbf{3.6\%}  & 6.3\%       \\
PF + Concat &  29.6 & 39.8 & 2.7\%  & \textbf{2.3\%}  & \textbf{40.7} & \textbf{46.6} & 3.9\%  & \textbf{1.5\%}  & 40.2 & 44.3 & 4.8\%  & \textbf{5.6\%}   \\ \hline
 \texttt{ChatGPT} \\
Concatenation         & 38.1   & 45.4   &   19.9\%  &   -   & 51.6   &  57.9  &   18.1\%      &  -            &  53.1     &  64.9  &   13.6\%      &       -       \\
Post-Fusion   & 40.1   &  50.4  &    7.4\%     &   12.0\%    &  51.4  &  57.3  &     9.1\%    &   10.2\%           &     57.1  &  71.2 &     2.1\%    &     4.3\%         \\ 
Pruning Prompt      &  39.0  & 50.5   &  6.9\%       &   -    & 52.7   &  59.5  &   8.1\%      &     -         &  47.7     & 62.6   &  6.7\%       &     -         \\
Summary Prompt     & 40.5   & 53.3   &   \textbf{5.1\%}     &   -      & 51.6   & 60.1   & 6.4\%        &    -          &  50.4     &  67.0  &      4.7\%   &   -           \\
Concat + PF  & 42.9   & 53.9   &   6.5\%       &   3.8\%  &  \textbf{55.9}       &  \textbf{62.8}  & 7.5\%   &  4.3\%                   &  60.6     & 74.0   &    \textbf{1.7\%}     &    2.2\%         \\
PF + Concat &  \textbf{43.2}  & \textbf{54.5}   &    5.4\%     &   \textbf{3.6\%}   &  54.0  & 61.7   &  
\textbf{6.2\%}       &  \textbf{3.1\%}           &  \textbf{63.9}     &  \textbf{76.9}  &   2.1\%      &  \textbf{2.0\%}     \\\hline
\texttt{GPT4} \\
Concatenation    & 41.9 & 52.9 & 14.9\% &  -     & 54.1 & 61.8 & 12.7\% & -    & 57.0 & 63.9 & 9.8\%  & -        \\
Post-Fusion   & 39.7 & 51.7 & \textbf{5.5\%}  & 13.4\% & 55.0 & 63.2 & 8.9\%  & 11.8\% & 58.2 & 64.5 & 3.5\%  & 6.7\%       \\ 
Pruning Prompt    &  41.2 & 52.3 & 6.2\%  & -      & 55.2 & 62.8 & \textbf{4.5\%}  & -      & 57.2 & 63.1 & 7.5\%  & -       \\
Summary Prompt    & 40.6 & 52.6 & 7.4\%  & -      & 54.8 & 62.5 & 5.9\%  & -      & 57.8 & 62.7 & 6.5\%  & -   \\
Concat + PF & \textbf{44.3} & \textbf{55.1} & 6.4\%  & \textbf{2.1\%}  & \textbf{58.3} & \textbf{67.4} & 7.1\%  & \textbf{3.2\%} & \textbf{66.2} & \textbf{78.4} & \textbf{3.8\%}  & \textbf{1.1\%}         \\
PF + Concat & 43.8 & 54.6 & 7.3\%  & 4.2\%  & 57.8 & 66.2 & 9.5\%  & 7.3\%  & 65.3 & 77.9 & 4.2\%  & 3.6\%     \\
\bottomrule
\end{tabular}
}
\caption{Exact match (EM) and F1 scores on filtered DEV split of the NQ, TriviaQA and SQuAD using Top-5 passages under with gold passage setting. \%Unk denotes the percentage of Unknown responses. \%NM denotes the error rate by majority vote. \textbf{Concat} refers to the Concatenation strategy and \textbf{PF} refers to Post-Fusion strategy.}
\label{tb_5psg_wg}
\end{table*}

\begin{table*}[ht]
\centering
\resizebox{0.9\linewidth}{!}{
\renewcommand{\arraystretch}{1}
\begin{tabular}{l|cccc|cccc|cccc}
\toprule
                & \multicolumn{4}{c}{\textbf{NQ}}           & \multicolumn{4}{c}{\textbf{TriviaQA}}     & \multicolumn{4}{c}{\textbf{SQuAD}}          \\ 
                & EM & F1 & \%Unk & \%NM & EM & F1 & \%Unk & \%NM & EM    & F1 & \%Unk & \%NM \\ \hline
Supervised     &  40.9  &  -  &     -    &    -   &  55.2  & -  &     -    &    -    &   35.8    &  -  &     -    &   -           \\ \hline \hline 
 \multicolumn{9}{l}{\textit{\textbf{Without gold passage}}}\\ \hdashline
 \texttt{LLama2} \\
Concatenation    & 24.6 & 34.6 & 18.2\% &  -     & 35.8 & 40.9 & 14.6\% &  -    & 20.1 & 28.9 & 21.8\% & - \\
Post-Fusion     & 24.9 & 36.3 & 13.8\% & 15.3\% & 35.9 & 43.8 & 10.5\% & 14.5\% & 21.5 & 29.5 & 16.2\% & 18.3\%       \\ 
Pruning Prompt      &  25.7 & 35.4 & 12.7\% & -     & 36.2 & 43.9 & 9.8\%  & -     & 23.5 & 30.4 & 10.4\% & -  \\
Summary Prompt    & 26.3 & 35.7 & 10.3\% & -     & 36.2 & 42.0 & 8.5\%  & -     & 23.8 & 30.2 & 10.9\% & -          \\
Concat + PF & \textbf{28.0} & \textbf{38.9} & \textbf{3.2\%}  & \textbf{3.6\%}  & 37.7 & 43.2 & \textbf{4.2\%}  & 3.5\%  & \textbf{26.5} & 34.9 & \textbf{3.2\%}  & 2.6\%       \\
PF + Concat & 27.9 & 38.5 & 8.7\%  & 4.8\%  & \textbf{38.2} & \textbf{43.6} & 8.9\%  & \textbf{2.8\%}  & 24.2 & \textbf{35.8} & 12.8\% & \textbf{2.3\%}     \\ \hline
\texttt{ChatGPT} \\
Concatenation    & 34.5   & 43.8   &   23.1\%  &   -   & 49.3   &  55.5  &   19.9\%      &  -            &  28.1     &  34.8  &   28.5\%      &       -       \\
Post-Fusion     & 38.3   &  48.3  &    10.1\%     &   9.0\%    &  49.7  &  55.7  &     10.7\%    &   7.4\%           &     32.1  &  40.3 &     13.9\%    &     12.3\%         \\ 
Pruning Prompt      &  36.2  & 46.3   &  9.1\%       &   -    & 49.3   &  56.5  &   9.5\%      &     -         &  36.1     & 40.6   &  12.7\%       &     -         \\
Summary Prompt    & 36.3   & 48.4   &    \textbf{8.6\%}     &   -      & 48.3   & 56.5   & \textbf{7.7\%}        &    -          &  34.1     &  40.0  &      13.7\%   &   -           \\
Concat + PF & \textbf{39.9}   & 49.7   &   9.3\%       &   5.3\%      &  \textbf{52.7}  & \textbf{59.5}   &   9.1\%  &  \textbf{2.8\%}       &  \textbf{40.1}     & \textbf{43.8}   &    \textbf{5.7\%}     &    \textbf{2.3\%}          \\
PF + Concat &  38.9  & \textbf{50.1}   &    9.1\%     &   \textbf{4.3\%}   &  50.5  & 57.7   &  6.7\%       &  3.2\%     &  38.5     &  41.2  &   9.9\%      &  5.4\%     \\ \hline
\texttt{GPT4} \\
Concatenation    & 36.9 & 50.6 & 18.9\% & -    & 51.3 & 60.7 & 16.7\% & -    & 29.7 & 30.9 & 25.8\% & -    \\
Post-Fusion     & 37.7 & 49.7 & 6.5\%  & 9.9\%  & 51.5 & 59.0 & 13.2\% & 8.9\%  & 33.1 & 37.8 & 12.8\% & 12.5\%\\ 
Pruning Prompt      &38.3 & 48.4 & 9.2\%  & -      & 51.2 & 58.2 & 12.5\% & -      & 32.7 & 39.8 & 13.6\% & -       \\
Summary Prompt    & 38.5 & 49.6 & 8.3\%  & -      & 50.8 & 58.5 & 13.9\% & -      & 35.9 & 39.2 & 12.5\% & -  \\
Concat + PF & \textbf{41.5} & \textbf{52.1} & \textbf{5.4\%}  & \textbf{3.1\%}  & \textbf{55.7} & \textbf{63.7} & \textbf{8.1\%}  & \textbf{3.8\%}    & 41.8 & 44.7 & \textbf{5.6\%}  & \textbf{3.2\%}         \\
PF + Concat & 40.6 & 51.6 & 6.9\%  & 9.2\%  & 54.3 & 62.8 & 12.5\% & 6.4\%  & \textbf{42.1} & \textbf{44.9} & 9.7\%  & 8.4\%    \\ 
\bottomrule
\end{tabular}
}
\caption{Exact match (EM) and F1 scores on filtered DEV split of the NQ, TriviaQA and SQuAD using Top-5 passages on without adding gold passage setting. \%Unk denotes the percentage of Unknown responses. \%NM denotes the error rate by majority vote. \textbf{Concat} refers to the Concatenation strategy and \textbf{PF} refers to Post-Fusion strategy.}
\label{tb_5psg}
\end{table*}

\textbf{Retriever and LLM model.}
We use the Wikipedia dump from Dec. 20, 2018 for NQ and TriviaQA and the dump from Dec. 21, 2016 for SQuAD. We apply preprocessing steps following~\citet{chen2017reading,karpukhin2020dense,liu2021dense}, which involve generating non-overlapping passages of 100 words each. Similar to \cite{izacard2021leveraging}, passages are retrieved with DPR~\cite{karpukhin2020dense} for NQ and TriviaQA and with BM25~\cite{robertson1995okapi} for SQuAD.
We consider two different settings for this study. The first utilizes the top-$k$ retrieved passages directly (gold passage is not necessarily included). In contrast, the second setting concerns the situation that the gold-standard passage is included in the context. If the gold passage is not within the top-$k$ passages, we randomly insert it into the top-$k$ list. 

We use both open and close LLMs. For Llama2~\cite{touvron2023llama}, we use the instruction-tuned version \texttt{Llama-2-7b-chat-hf} model and apply greedy decoding with the temperature parameter set to 0. For ChatGPT, we use the \texttt{gpt-3.5-turbo-16k} model. For GPT4~\cite{openai2023gpt4}, our choice is~\texttt{gpt-4-0613}.

\subsection{Results} \label{main_result}
The results using the gold passages setting are presented in Table~\ref{tb_5psg_wg}, while those without incorporating gold passages are in Table~\ref{tb_5psg}. Initially, we obtain the Top-5 retrieved passages, representing the setting without added gold passages. If these passages don't contain the answer, we randomly integrate the gold passage among the Top-5 candidate passages, corresponding to the setting with gold passages. 

Table~\ref{tb_5psg_wg} reveals that among the single-round zero-shot methods, Post-Fusion consistently surpasses the traditional concatenation approach in both EM and F1 metrics across all three benchmarks. This indicates that the model may become distracted when faced with a combination of relevant passages.
Compared to zero-shot and few-shot approaches, both Pruning Prompt and Summary Prompt show a marked enhancement over the concatenation method, though the margin of improvement is modest.  
The use of the CoT, which elicits a potential reasoning process, can guide the model in attending to relevant passages. However, this approach does not greatly enhance single-hop question answering as compared to prior multi-hop reasoning studies~\cite{wei2022chain,trivedi2022interleaving}. 

Compared to single-round methods, multi-round strategies consistently deliver superior performance, showcasing significant improvements. For instance, on the NQ dataset, Concat + PF exceeds the Concatenation method by over $10\%$ on average across three distinct LLMs. 
It suggests the efficacy of integrating model uncertainty as feedback. Among the multi-round approaches, Concat + PF demonstrates better performance compared to PF + Concat on most of cases. Comparing PF + Concat with Post-Fusion, it is evident that PF + Concat, leveraging LLM to select the best answer from a candidate pool, outperforms the majority vote approach. 

In the realm of open-domain question-answering, as evidenced by Table \ref{tb_5psg}, the performance metrics (EM and F1) under settings without the addition of a gold passage are comparatively lower. This is primarily attributed to the reduced recall of Top-k retrieval, resulting in a higher propensity to generate ``unknown'' responses. Notably, our proposed multi-round methodologies, when leveraging GPT4 as the LLM, deliver performance figures that are on par with supervised outcomes.

\begin{figure}[t]
\centering
\includegraphics[width=0.9\linewidth]{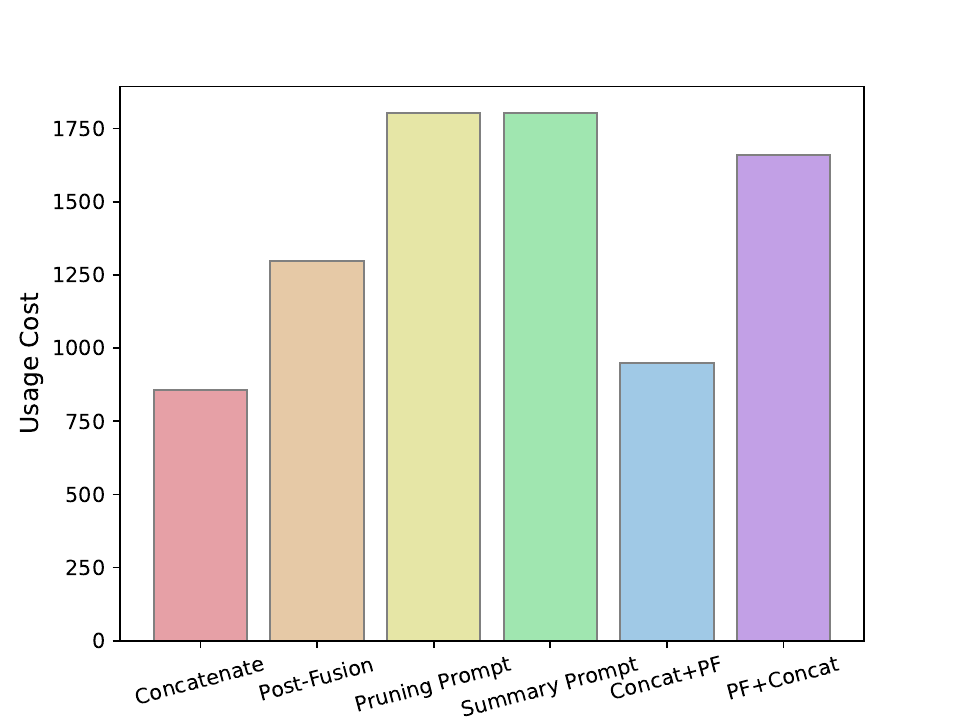}
\caption{The token usage of different approaches using top-5 passages.}
\label{figure:nq_usage}
\end{figure}

\subsection{Usage Analysis}
Striking a balance between enhancing the quality of generated answers and optimizing resource utilization is essential. As depicted in Figure~\ref{figure:nq_usage}, different methodologies vary in their token usage. The Concatenate method is the most resource-efficient, whereas the Concat + PF method, albeit being the second most efficient, has an additional 90.5 tokens on average when compared to Concatenate. Given the significant performance boost of Concat + PF over Concatenate (a $15.6\%$ increase in EM as presented in Table \ref{tb_5psg}), we advocate for the adoption of Concat + PF. This offers a more efficient means of integrating retrieved passages with LLMs.

\subsection{Effect of different Top-k passages from the retriever}
\begin{figure}[t]
\centering
\includegraphics[width=0.9\linewidth]{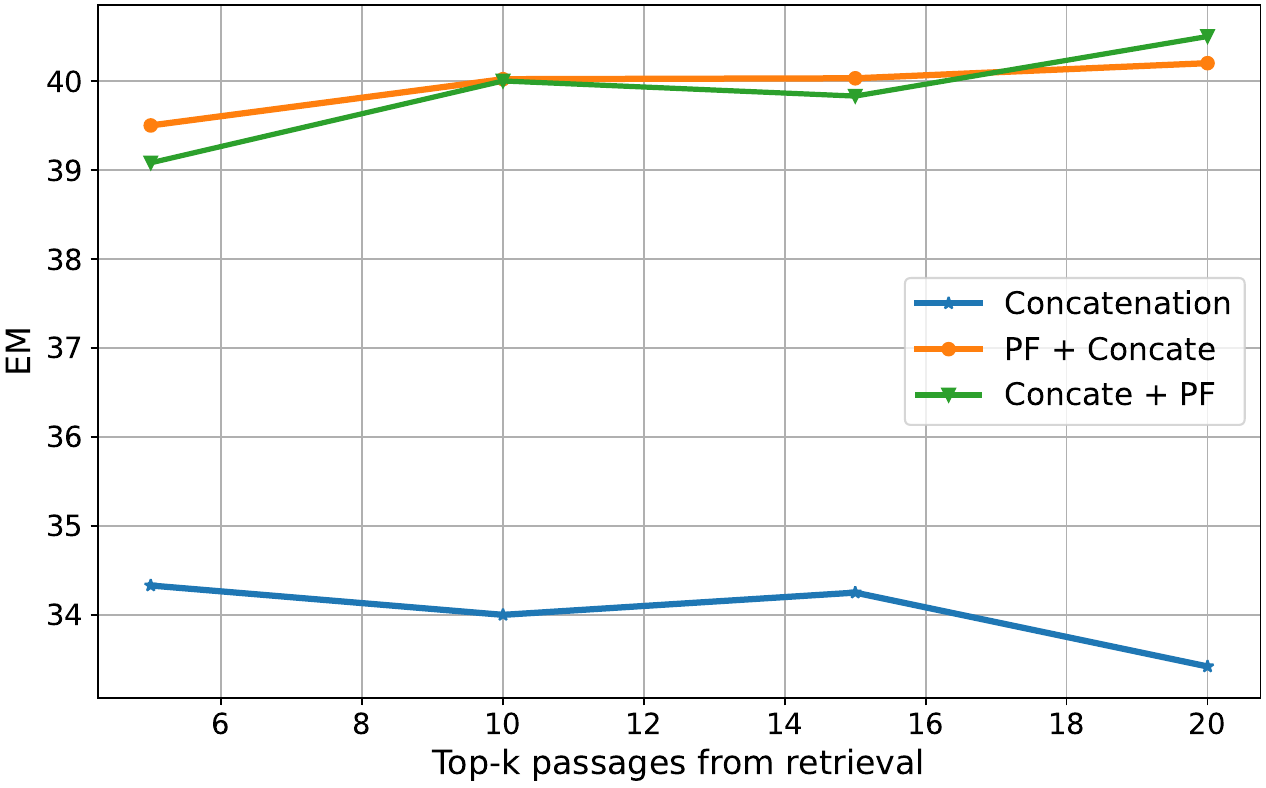}
\caption{The answer EM performance with the increase of Top-k retrieved passages.}
\label{figure:nq_retrieve}
\end{figure}

Figure \ref{figure:nq_retrieve} showcases open-domain QA results using the Top-k retrieved passages on NQ dataset. As k increases, we observe a corresponding increase in retrieval recall. Our multi-stage methods, Concat + PF and PF + Concat, both benefit from increasing k values, showing enhancements of 1.5 and 0.7 points, respectively, when moving from Top 5 to 20. In contrast, the conventional concatenation method experiences a 0.8 EM performance decline from Top 5 to 20. This suggests that the concatenation prompt can become counterproductive with the inclusion of more passages, potentially because it struggles to identify the correct passage and gets distraction by incorrect ones. However, our multi-stage approaches remain undeterred with the addition of passages, demonstrating greater robustness.

\subsection{Effect of different Decoding Strategies}
Instead of the traditional greedy decoding strategy, a newer method known as self-consistency \cite{wang2022self} has been introduced and employed in the chain-of-thought prompting \cite{wei2022chain}. This method begins by sampling from the language model's decoder to produce a diverse set of answers. The optimal answer is then obtained by marginalizing the samples' reasoning paths. 

For the concatenation prompt, we opt for temperature sampling \cite{ackley1985learning,ficler2017controlling} as our decoding strategy, yielding $p$ outputs, rather than generating a singular answer via greedy decoding as detailed in section \ref{main_result}. In the case of the post-fusion prompt, each passage employs a sampling decoding strategy, generating $p$ outputs for every $k$ passages. This results in a total of $p \times k$ outputs. It's important to distinguish between post-fusion prompts and self-consistency. The former pertains to using different inputs, while the latter is about the decoding sampling strategy.

\begin{figure}[t]
\centering
\includegraphics[width=0.9\linewidth]{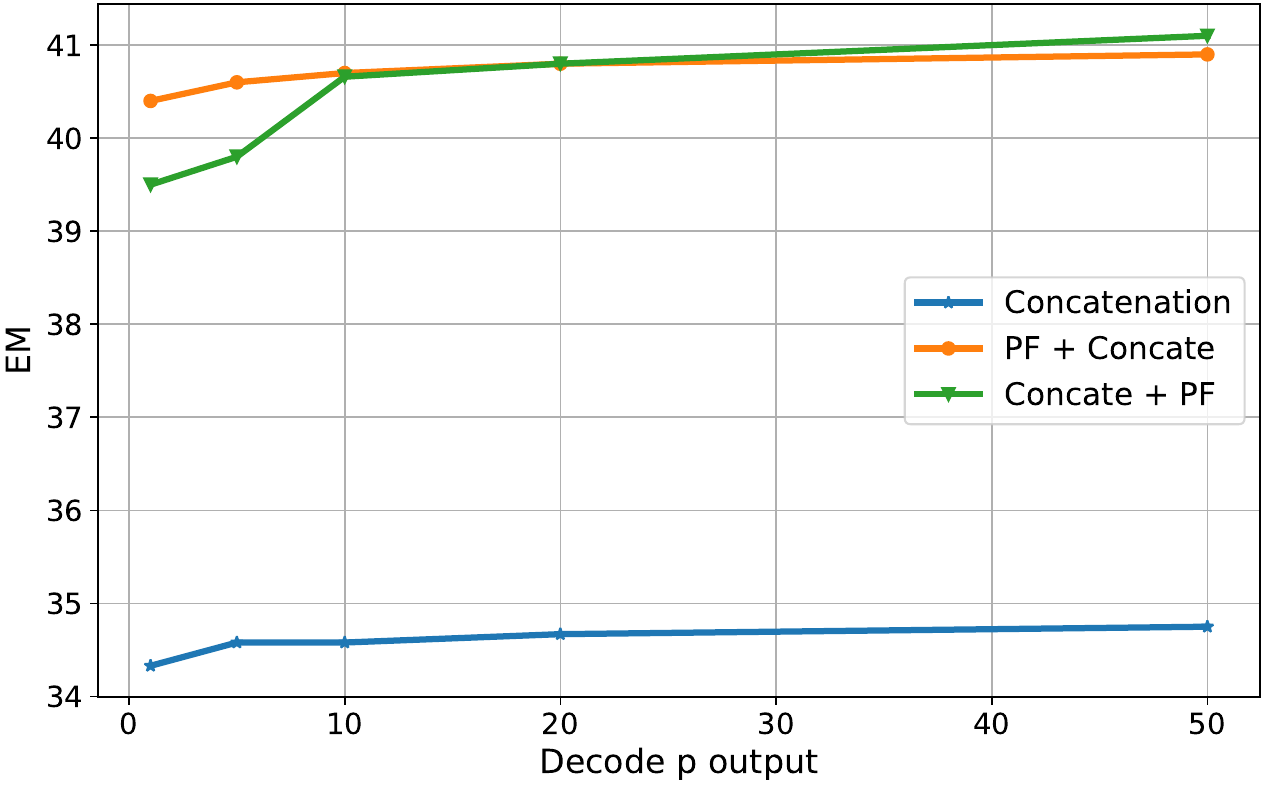}
\caption{The answer EM performance with the increase of the number of decode output.}
\label{figure:nq_decode}
\end{figure}

Figure \ref{figure:nq_decode} presents an ablation of results with a temperature of 0.7 and varying values of $p$ in Top-$p$ sampling on ChatGPT, using the Top-5 retrieved passages from the NQ dataset. The data suggests that small sampling outputs, ranging from 1 to 10, significantly enhance performance. However, as $p$ increases from 10 to 50, the degree of improvement diminishes. And Concate + PF approach could benefit more from the increase of $p$.

\subsection{Effect of the order of the gold passage}
In this section, we aim to assess how the placement of the gold passage within the Top-$k$ passages influences the ability of the LLM to generate accurate answers. We examine three different placements: (1) consistently positioning the gold passage at the start of the Top-$k$ passage list; (2) consistently placing the gold passage at the end of the Top-$k$ passage list; (3) maintaining the original sequence produced by the retrieval model.

\begin{figure}[t]
\centering
\includegraphics[width=0.9\linewidth]{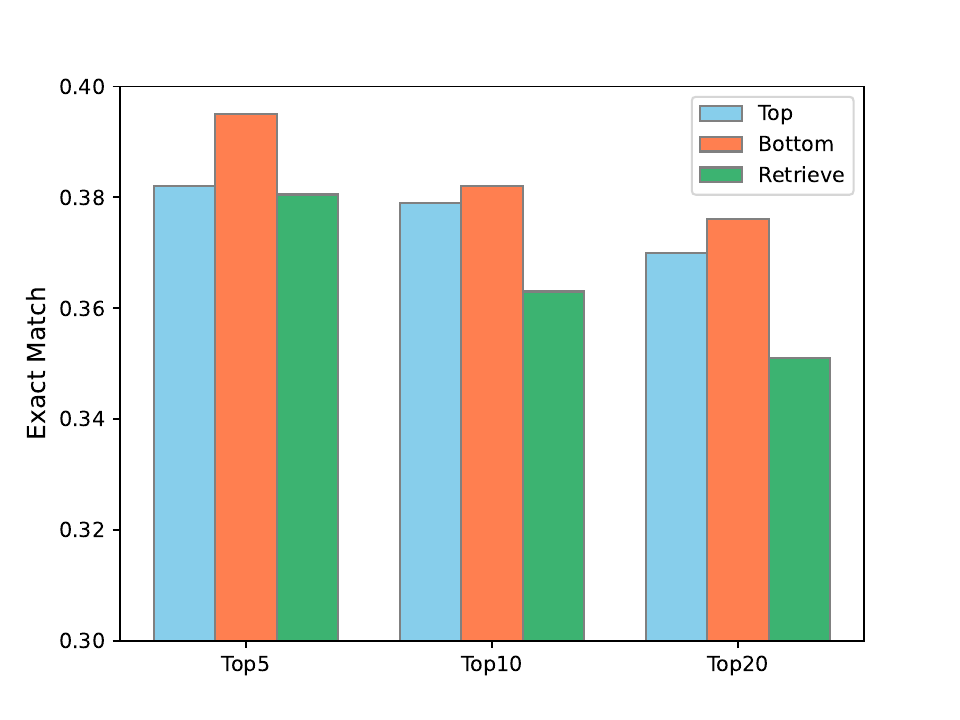}
\caption{The impact on the position of gold passage on Combination method.}
\label{figure:nq_order}
\end{figure}

As the results depicted in Fig. \ref{figure:nq_order}, it is evident that the placement of the gold passage significantly affects the quality of the generated answers. Consistently placing the gold passage in the same position tends to improve performance compared to using the retrieval order. Among the constant placement options, positioning the gold passage at the bottom tends to yield better results than placing it at the top. This outcome might be tied to our prompt design, where we present the Top-$k$ passages first, followed by the question. Consequently, keeping the gold passage closer to the question seems to enhance performance to the greatest extent. Moreover, this observation is aligned with the \cite{liu2023lost}, where they find that a distinctive U-shaped performance, as performance peaks when key information is at the start or end of the input, but drops significantly for mid-context information.

\section{Related Work}

The recent proliferation of LLM-powered applications, such as ChatGPT/GPT4~\cite{openai2023gpt4}, Bing Chat, and CoPilot, has highlighted both the impressive performance and certain limitations of LLMs. These limitations include a high compute and data demand, making it a challenge to continually update LLMs both efficiently and effectively~\cite{scialom2022continual}. LLMs also tend to generate plausible yet non-factual texts, a phenomenon known as ``hallucination''~\cite{openai2023gpt4,zhao2023knn}. In response to these issues, the field is witnessing a trend towards augmenting LLMs with specialized tools~\cite{schick2023toolformer,paranjape2023art}, such as code interpreters~\cite{zhang2021disentangled,gao2022pal,shao2023synthetic} or search engines~\cite{park2023query}. The goal is to delegate specific tasks to more proficient systems or to enrich the LLMs' input context with more pertinent information.

Augmentation of language models with pertinent data retrieved from diverse knowledge bases has demonstrated its effectiveness in enhancing open-domain question answering performance~\cite{lazaridou2022internet,izacard2022few,chen2022augmenting}. The process typically involves using the input query to (1) command a retriever to fetch a document set (essentially, token sequences) from a corpus, after which (2) the language model integrates these retrieved documents as supplemental information, guiding the final prediction.

The interleaving between the retriever and LLM could be considered a reciprocal process. Various studies have been conducted on generation-augmented retrieval (GAR), which involves revising or supplementing queries with generated background information to enhance the retrieval of relevant content. Well-known examples of this approach include GAR~\cite{mao2020generation} and HyDE~\cite{gao2022precise}. With regard to complex multi-step reasoning questions, work involving LLMs often necessitates the retrieval of segmented knowledge~\cite{trivedi2022interleaving,khattab2022demonstrate}. This chain-of-thought reasoning process~\cite{wei2022chain,jiang2023active,nguyen2023cof} is followed by conducting partial reasoning to generate the next question, then retrieving further information based on the outcome of that partially formed next question, and repeating this cycle as needed~\cite{yao2022react,press2022measuring}.

Our work primarily focuses on a specific scope: once the output from the retriever is determined, we aim to identify the most effective method of inputting this data into LLMs for answer generation.

\section{Conclusion}
In this study, we identified two key challenges associated with integrating LLMs and retrieved passages: the occurrence of ``unknown'' responses when feeding LLMs with concatenated passages and the erroneous majority when using the Post-Fusion approach.
To overcome these challenges, we proposed four improved approaches, including two CoT-related strategies and two multi-round methods incorporating LLM's feedback. Through our experimental results and token usage analysis, we observed that it is advantageous to first employ a concatenation strategy to generate an answer. In the case of an ``unknown'' response, we recommend transitioning to the Post-Fusion approach to obtain the final answer through a majority vote.

\section*{Limitations}
Our evaluation is primarily constrained to three open-domain QA datasets to align better with the supervised state-of-the-art approach cited in \cite{izacard2021leveraging}. To ensure the broader applicability and robustness of our findings, it's essential to evaluate the proposed methods on other benchmarks, including MS MARCO and WebQuestions datasets~\cite{nguyen2016ms,berant2013semantic}.

Currently, our evaluation focuses predominantly on textual QA. While the proposed approach seems generalizable to other modalities like tables~\cite{pasupat2015compositional,zhu2021tat} and knowledge bases~\cite{berant2013semantic,bao2016constraint}, we have yet to empirically test and validate this claim. Future studies could delve into exploring its effectiveness on diverse modalities like UniK QA~\cite{oguz2022unik}.

We haven't thoroughly evaluated how our approach scales with larger datasets or more complex queries~\cite{trivedi2022musique}. This could be an avenue of exploration, as scalability is vital for real-world applications.

% To have better comparison with supervised SOTA approach \cite{izacard2021leveraging}, we only evaluate on three open-domain QA datasets. To further test our findings and the efficacy of proposed methods, we plan to evaluate them on additional open-domain single-hop question answering benchmarks, such as the MS MARCO, WebQuestions datasets~\cite{nguyen2016ms,berant2013semantic}.
% Moreover, current evaluation is on the textual QA, our approach is also general to table~\cite{zhu2021tat} and knowledge base QA, in future, we also want to verify our findings on the other modality QA~\cite{oguz2022unik}.

\appendix
\clearpage

\label{sec:appendix}

% \textbf{Scaling with number of passages.}
% In Fig. \ref{fig:scale_with_k}, we report the performance with respect to the number of retrieved passages.
% \begin{figure}
%     \centering
%     \begin{minipage}[b]{0.5\textwidth}
%         \includegraphics[width=\textwidth]{Figure/nq_direct_topk.pdf}
%         \label{fig:first_image}
%     \end{minipage}
%      \hfill
%     \begin{minipage}[b]{0.5\textwidth}
%         \includegraphics[width=\textwidth]{Figure/nq_seperate_topk.pdf}
%         \label{fig:second_image}
%     \end{minipage}
%         \caption{Top figure shows using combination the EM and the percentage of Unknown with the different k passages; Bottom figure shows using separate way, the EM, the percentage of Unknown and Not Majority on Natural Question Dataset.}
%         \label{fig:scale_with_k}
% \end{figure}
\section{Prompt used in Different Approaches} \label{prompt_approach}
\begin{figure*}[t]
\centering
\includegraphics[width=\linewidth]{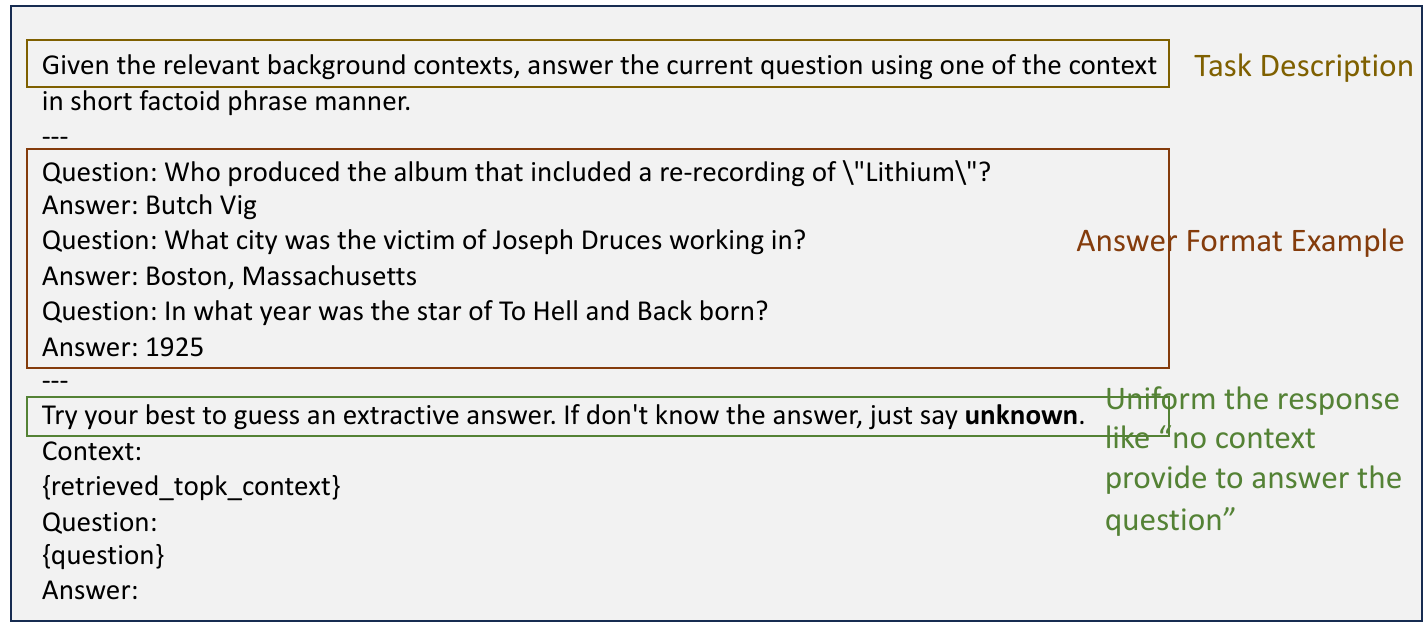}
\caption{The Prompt used in Concatenation and Post-Fusion.}
\label{figure:direct}
\end{figure*}

The prompts used in the Concatenation and Post-Fusion approaches are illustrated in Fig. \ref{figure:direct}. In the Concatenation approach, \texttt{retrieved\_topk\_context} represents the concatenation of the top-k retrieved passages. Conversely, in the Post-Fusion approach, it represents a single passage at a time.

\begin{figure*}[t]
\centering
\includegraphics[width=\linewidth]{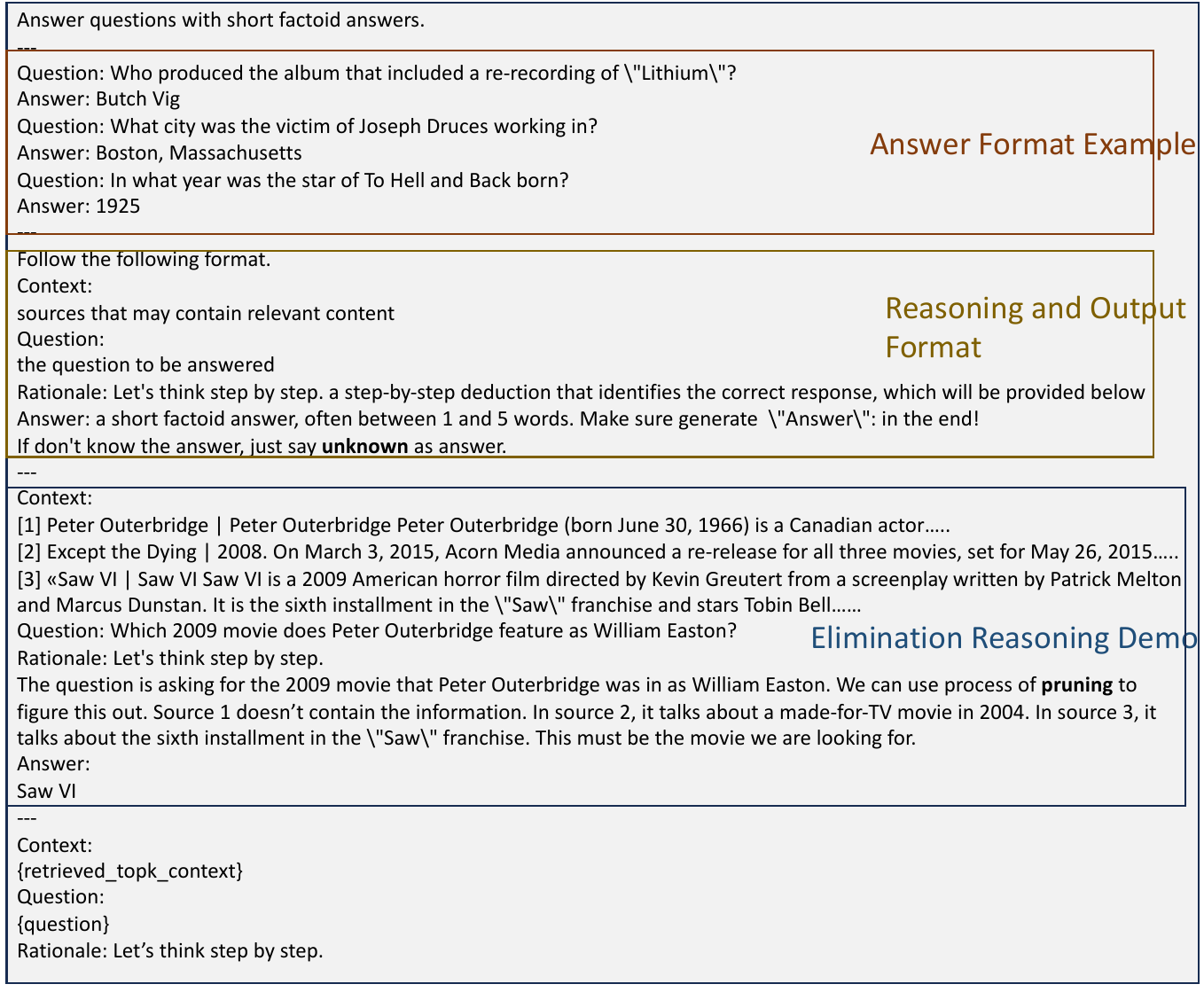}
\caption{The Pruning Prompt.}
\label{figure:pruning}
\end{figure*}

\begin{figure*}[t]
\centering
\includegraphics[width=\linewidth]{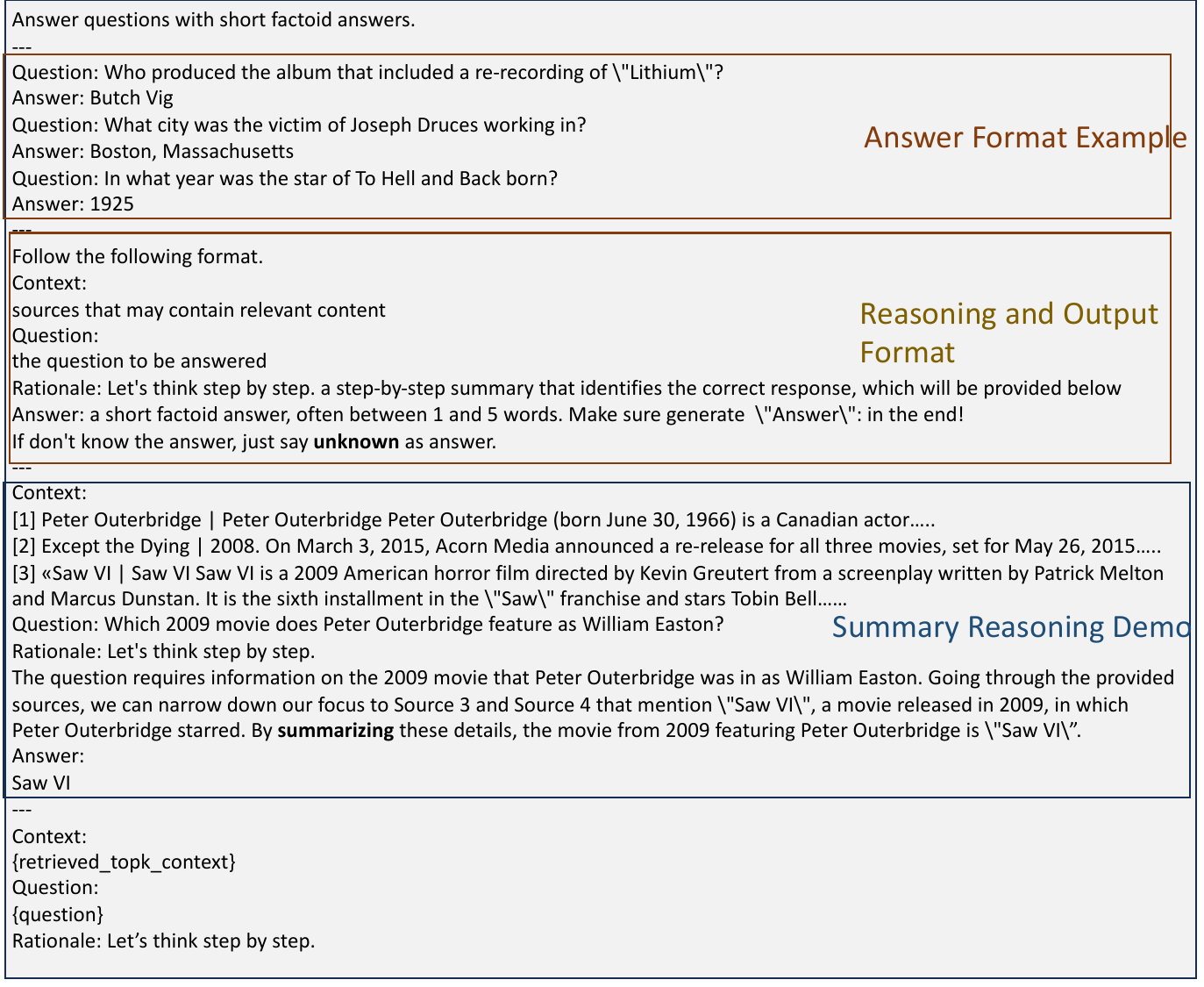}
\caption{The Summary Prompt.}
\label{figure:summary}
\end{figure*}

The Pruning Prompt's specific prompt is presented in Fig. \ref{figure:pruning}, while the Summary Prompt's prompt is depicted in Fig. \ref{figure:summary}.

\end{document}